\documentclass[%
preprint%
,secnumarabic%
,tightenlines%
,amssymb, amsmath,nobibnotes, aps,nofootinbib,showpacs,]{revtex4}
\usepackage{epsfig}%
\usepackage{graphicx}%
\expandafter\ifx\csname package@font\endcsname\relax\else
 \expandafter\expandafter
 \expandafter\usepackage
 \expandafter\expandafter
 \expandafter{\csname package@font\endcsname}%
\fi
\begin{document}

\title{Consequences of $f(R)$ theories of gravity on gravitational leptogenesis}

\author{G. Lambiase$^{a,b}$, S. Mohanty$^c$, and L. Pizza$^{a,d}$}
\affiliation{$^a$University of Salerno, 84084 - Fisciano (SA), Italy.\\
$^b$INFN, Sezione di Napoli, Italy,}
\affiliation{$^c$Physical Research Laboratory, Ahmedabad 380009, India.}
\affiliation{$^d$Università di Pisa,  56126 - Pisa, Italy}

\def\be{\begin{equation}}
\def\ee{\end{equation}}
\def\al{\alpha}
\def\bea{\begin{eqnarray}}
\def\eea{\end{eqnarray}}

\renewcommand{\theequation}{\thesection.\arabic{equation}}

\begin{abstract}
$f(R)$-theories of gravity are reviewed in the framework of the matter-antimatter asymmetry in the Universe.
The asymmetry is generated by the gravitational coupling of heavy (Majorana) neutrinos with the Ricci scalar curvature.
In order that the mechanism works, a time varying non-zero Ricci curvature is necessary.
The latter is provided by $f(R)$ cosmology, whose Lagrangian density is of the form ${\cal L}(R)\sim f(R)$.
In particular we study the cases $f(R)\sim R+\alpha R^n$ and $f(R)\sim R^{1+\epsilon}$.

\end{abstract}

\maketitle

\section{Introduction}


The observation that the present phase of the expanding Universe is accelerated \cite{accUn} has motivated in the last years
the developments of many models of gravity which go beyond the general relativity, and therefore the standard
cosmological model. Among the different approaches, the $f(R)$-theories of gravity have received a great attention.
The reason relies on the fact that they allow to explain, via a gravitational dynamics, the
observed accelerating phase of the Universe, without invoking exotic matter as sources of dark energy.
Moreover, they also provide an alternative approach Dark Matter problem.

The Lagrangian density of these  models  does depend on higher-order curvature invariants (such as, for example, $R^2$, $R_{\mu\nu}R^{\mu\nu}$,
$R\Box R$, and so on) \cite{faraonibook,defelice}.
In this paper we focalize our attention to $f(R)$ models which are a generic function of the Ricci scalar
curvature $R$
\begin{equation}\label{Lagr}
  S=\frac{1}{2\kappa^2}\int d^4x \sqrt{-g}\, f(R)+S_m[g_{\mu\nu},\psi]\,.
\end{equation}
In particular, the function $f(R)$ we concern are of the forms: $f(R)=R+\alpha R^n$ ($n=2$) and $f(R)=R^m$, with $m=1+\epsilon\sim {\cal O}(1)$ ($\epsilon \ll 1$).
In Eq. (\ref{Lagr}), $S_m$ is the action of matter and $\kappa^2=8\pi G=8\pi m_P^{-2}$ ($m_P\simeq 10^{19}$GeV is the Planck mass).
Cosmological and astrophysics consequences of (\ref{Lagr}) have been largely studied in literature \cite{allJCAP} (see also \cite{faraoniRMP,faraonibook,defelice,nojiri,capozziello-f}).

The aim of this paper is to study the origin of the baryon number asymmetry in the framework of $f(R)$ theories of gravity.
As well known, the matter-antimatter asymmetry in the Universe is still an open problem of
the particle physics and cosmology \cite{kolb}.
The successful prediction of the big bang nucleosynthesis (BBN)
\cite{Copi,burles} and the observations of CMB anisotropies combined with the large structure of the
Universe \cite{wmap,bennet}
show that the baryon to photon number of the universe, i.e. the parameter that characterize such a a asymmetry,  is of the order
 \begin{equation}\label{etaexp}
\eta_{\mbox{exp}} \equiv \frac{n_B-n_{\bar B}}{s}\lesssim (9.2 \pm 0.5)\,\, 10^{-11}\,,
 \end{equation}
where $n_B$ ($n_{\bar B}$) is the baryon (antibaryon) number density, and $s=(2\pi^2/45)g_* T^3$ is the entropy density ($g_*\simeq 100$
are the relativistic degrees of freedom).

As shown by Sakharov, a (CPT invariant) theory is able to explain the baryon asymmetry provided that the following conditions (Sakharov's conditions) are fulfilled \cite{sakharov}: 1) there must exist processes that violate the baryon number; 2) the discrete symmetries
C and CP must be violated; 3) departure from thermal equilibrium.
However, as shown in \cite{cohen},  a dynamical violation of CPT (which implies a different spectrum of particles and antiparticles)
may give rise to the baryon number asymmetry also in a regime of thermal equilibrium.
A successful mechanism for explaining the asymmetry between matter and anti-matter is provided by Leptogenesis \cite{fukugita}. In this scenario,
in which the Majorana neutrino is introduced, it is possible to generate the baryon asymmetry if the asymmetry is generated in the lepton sector
at either GUT or intermediate scales, even if the baryon number is conserved at high energy scales.

In the present paper we show that the coupling of the heavy neutrinos with the gravitational background gives rise to an effective potential that modifies differently the dispersion relations of neutrinos with left and right helicity, leading to a net lepton asymmetry even if neutrino are in thermal equilibrium\footnote{The coupling of the chiral fields  to the cosmological background is odd under $CP$  which biases the generation of leptons over anti-leptons in the presence of lepton number violating interactions at  thermal equilibrium (similar to \cite{cohen}). The thermodynamic interpretation of such a scenario is described, as usual \cite{cohen,kitano}, by the introduction of an {\it effective} chemical potential which is different for particles and antiparticles owing to the CP odd coupling.}. The lepton asymmetry is then converted into baryon asymmetry by the action of sphalerons in the elctroweak era.
For this mechanism to work, a time varying non-zero Ricci curvature is required during the radiation dominated era. This is provided by $f(R)$ cosmology
(mechanisms for baryo/leptogenesis, similar to one discussed in this paper, can be found in \cite{kitano}).

The paper is organized as follows. In Section II we discuss the leptogenesis mechanism based on the curvature coupling of heavy neutrinos.
In Section III we derive the main equations of $f(R)$ models of gravity, studying also the constraints provided by BBN physics.
In Section IV and V we investigate the baryon asymmetry for some specific models of $f(R)$.
Conclusions are shortly discussed in Section VI.

\section{CP odd gravitational coupling of Majorana neutrinos}

In this Section, we study the generalization in the matter Lagrangian by including higher order terms in $R$ consistent with general covariance, Lorentz-invariance in a locally inertial frame. Consider the action for a four component Dirac fermion
$\psi$ which couples to background gravity,
\begin{eqnarray}
S_m[g_{\mu \nu},\psi] &=& \int d^4x \sqrt{-g} \left[i\bar \psi \gamma^\mu (\overrightarrow\partial_\mu-\overleftarrow\partial_\mu) \psi -\right. \label{hR} \\
 & &  \left. - h_1(R)\, \bar \psi \psi - i h_2(R)\, \bar \psi \gamma_5 \psi \right]\,, \nonumber
 \end{eqnarray}
where $h_1(R)$ and $h_2(R)$ real valued scalar functions of the curvature,
 \bea
 h_1(R)&=& M + g_1(R)\,,  \nonumber\\
 h_2(R)&=& M^\prime + g_2( R)\,.
 \eea
Here $h_1$ is a generalization of the neutrino mass term. Note that since $\bar \psi \gamma_5 \psi$ transforms as a pseudo-scalar, the $h_2$ term is odd under $CP$.
 We write the four-component fermion
 \be
 \psi=\left(
        \begin{array}{c}
          \psi_L \\
          \psi_R \\
        \end{array}
      \right)\,.
 \ee
The lagrangian in terms of the two-component fields $\psi_R$ and $\psi_L$ becomes
 \bea
  {\cal L}&=&i \psi_R^\dagger\, \bar \sigma^\mu (\overrightarrow\partial_\mu-\overleftarrow\partial_\mu) \psi_R
  +i\psi_L^\dagger\,  \sigma^\mu (\overrightarrow\partial_\mu-\overleftarrow\partial_\mu)   \psi_L - \nonumber\\
  &-&\,h_1(\psi_R^\dagger \psi_L +\psi_L^\dagger \psi_R)-i\,h_2(\psi_R^\dagger \psi_L-\psi_L^\dagger \psi_R)\,,
 \label{L1}
 \eea
where $\sigma^\mu=(I, \sigma^i)$ and $\bar \sigma^\mu=(I,-\sigma^i)$ in terms of the Pauli matrices. The $h_2$ term can be rotated away by a chiral transformation
 \be
 \psi_L \rightarrow e^{-i \alpha/2} \psi_L \quad  \psi_R \rightarrow e^{i \alpha/2} \psi_R\,.
\ee
Keeping terms to the linear order in $\alpha$, we see that the lagrangian (\ref{L1}) changes by the amount
\bea
\delta {\cal L}&=& - \psi_R^\dagger \psi_R \bar \sigma^\mu \partial_\mu \alpha + \psi_L^\dagger \psi_L  \sigma^\mu \partial_\mu \alpha - \nonumber\\
&-&  h_1 ( i \alpha) \left( \psi_L^\dagger \psi_R - \psi_R^\dagger \psi_L \right) -i h_2 ( i \alpha)\left( \psi_L^\dagger \psi_R + \psi_R^\dagger \psi_L \right)\,.
\label{dL1}
\eea
Now we choose $\alpha=-h_2/h_1$ to eliminate the chiral mass term and obtain for the total Lagrangian
\bea
{\cal L}&=&i\psi_R^\dagger\, \bar \sigma^\mu (\overrightarrow\partial_\mu-\overleftarrow\partial_\mu) \psi_R
  +i\psi_L^\dagger\,  \sigma^\mu (\overrightarrow\partial_\mu-\overleftarrow\partial_\mu) \psi_L - \nonumber\\
  &-& \psi_L^\dagger \psi_L \sigma^\mu \partial_\mu \left(\frac{h_2}{ h_1}\right)+ \psi_R^\dagger \psi_R \bar \sigma^\mu \partial_\mu \left(\frac{h_2}{ h_1}\right)-\nonumber\\
  &-& \frac{1}{h_1} (h_1^2 + h_2^2) \left( \psi_L^\dagger \psi_R + \psi_R^\dagger \psi_L \right)\,.
 \label{l2a}
 \eea
If $h_1$ and $h_2$ are constants then, one can always rotate the axial-mass term away. We will assume that the neutrino mass $M \gg g_1$ therefore $h_1 \simeq M$ and since a constant $M^\prime$ can be rotated away $h_2= g_2$. Further we will assume that the background curvature is  only dependent on time. The lagrangian (\ref{l2a}) then reduces to the form
\bea
{\cal L}&=&i \psi_R^\dagger\, \bar \sigma^\mu (\overrightarrow\partial_\mu-\overleftarrow\partial_\mu) \psi_R
  +i \psi_L^\dagger\,  \sigma^\mu (\overrightarrow\partial_\mu-\overleftarrow\partial_\mu) \psi_L - \nonumber\\
  &-& \psi_L^\dagger \psi_L  \left(\frac{\dot g_2}{M}\right)+ \psi_R^\dagger \psi_R \left(\frac{\dot g_2}{M}\right) - \nonumber\\
  &-&  M \left( \psi_L^\dagger \psi_R + \psi_R^\dagger \psi_L \right)\,.
 \label{l2}
 \eea
The equation of motion for the left and the right helicity fermions derived from (\ref{l2}) are
\bea
 i\bar \sigma^\mu {\partial_\mu} \psi_R
 + \left(\frac{\dot g_2}{M}\right)  \psi_R- M \psi_L =0\,,  \nonumber\\
i \sigma^\mu {\partial_\mu} \psi_L
 - \left(\frac{\dot g_2}{M}\right) \psi_L - M   \psi_R =0\,.
 \label{eom1}
 \eea
Written in momentum space $\psi(x)=\psi(p) e^{i (Et -\vec p \cdot \vec x)}$ the equation of motion of $\psi_R$ and $\psi_L$ are
 \bea
 \left(E_R -\frac{\dot g_2}{M}\right)  \psi_R-  \vec \sigma \cdot \vec p \psi_R - M   \psi_L =0\,, \nonumber\\
 \left(E_L +\frac{\dot g_2}{M}\right)  \psi_L+  \vec \sigma \cdot \vec p \psi_L - M   \psi_R =0\,.
 \label{eom2}
 \eea
The canonical momenta of the $\psi_L$ and $\psi_R$ fields are as usual
\be
\pi_L=\frac{\partial {\cal L}}{\partial \dot \psi_L}=i \psi_L^\dagger,
\quad \pi_R=\frac{\partial {\cal L}}{\partial \dot \psi_R}=i \psi_R^\dagger\,.
\ee
Then the canonical Hamiltonian density is
\bea
{\cal H}&\equiv& \pi_L \dot \psi_L +\pi_R \dot \psi_R -{\cal L}\nonumber\\
&=&i \psi_L^\dagger \dot \psi_L +i \psi_L^\dagger{\bf \sigma}\cdot {\bf \nabla}\psi_L +i \psi_R^\dagger \dot \psi_R
-i \psi_R^\dagger {\bf \sigma}\cdot {\bf \nabla}\psi_R + M  \left( \psi_L^\dagger \psi_R + \psi_R^\dagger \psi_L \right)\nonumber\\
& & +n_L \left(\frac{\dot g_2}{M}\right)-n_R \left(\frac{\dot g_2}{M}\right)\,,
\eea
where we have introduced the number density operators of the left and right chirality modes,
\be
n_L\equiv  \psi_L^\dagger \psi_L , \quad n_R \equiv \psi_R^\dagger \psi_R\,.
\ee
The partition function in terms of this effective Hamiltonian is
\be
{\cal Z}= Tr e^{-\beta {\cal H}}\equiv Tr e^{-\beta( {{\cal H}_0-\mu_L n_L-n_R \mu_R})}\,,
\ee
where $\beta=1/T$ and ${\cal H}_0$ is the free particle Hamiltonian. We see
that when $\dot g_2$ is non-zero then the {\it effective} chemical potential for the left chirality neutrinos is $\mu_L=-\dot g_2/M$ and for the right-chirality neutrinos is $\mu_R= \dot g_2/M$.
In the presence of interactions which change $\psi_L \leftrightarrow \psi_R$ at thermal equilibrium there will be a net difference between the left and the right chirality particles,
\bea
n_R-n_L&=&\frac{1}{\pi^2}\int d^3 p \left[\frac{1}{1+e^{\beta({E_p-\mu_R})}}-\frac{1}{1+e^{\beta({E_p-\mu_L})}}\right]\nonumber\\
&=& \frac{T^2}{3}\frac{\dot g_2}{M}\,.
\eea
In this paper we consider the simplest case of $h_2(R)=g_2(R)=R/m_P$ as a linear function of the curvature $R$. The axial term in (\ref{hR}) is a
CP violating interaction between fermions and the Ricci curvature described by the dimension-five
operator \cite{lambiaseJCAP,lambiasePRD2011}
\begin{equation}
{\cal L}_{\diagup{\!\!\!\!\!\!C\!\!P} }= \sqrt{-g}\,\frac{1}{m_P} \, R
\bar \psi \,i\gamma_5 \psi\,. \label{cpv0}
\end{equation}
This operator is invariant under Local Lorentz transformation and
is even under $C$ and odd under $P$ and conserves $CPT$.
In a non-zero background $R$, there
is an effective $CPT$ violation for the fermions. Take $\psi=(N_R, N_R^c)^T$, where $N_R$ is a heavy right handed neutrino and $N_R^c$ a left handed heavy neutrino, which decay into the light neutrinos. Majorana neutrino interactions with the light neutrinos and Higgs  relevant for leptogenesis, are
described by the lagrangian
 \begin{equation}
 {\cal{L}}=- h_{\alpha \beta}(\tilde{\phi^\dagger}~
 \overline{N_{R \alpha}} l_{L \beta})
-\frac{1}{2}N_R^c \,{\widetilde M}\,N_R +h.c. \,,
\label{LN}
 \end{equation}
where ${\widetilde M}$ is the right handed neutrino mass-matrix, $l_{L \alpha}=
(\nu_\alpha , e^-_\alpha)_L^T $ is the left-handed lepton doublet
($\alpha $ denotes the generation), $\phi=(\phi^+,\phi^0)^T $ is
the Higgs doublet. In the scenario of leptogenesis introduced by
Fukugita and Yanagida, lepton number violation is
achieved by the decays $N_R \rightarrow \phi + l_L$ and also
${N_R}^c \rightarrow \phi^\dagger + {l_L}^c$. The difference in
the production rate of $l_L$ compared to $l_L^c$, which is
necessary for leptogenesis, is achieved via the $CP$ violation. In
the standard scenario, $n(N_R)=n(N_R^c)$ as demanded by $CPT$, but
$\Gamma(N_R \rightarrow l_L + \phi) \not = \Gamma(N_R^c
\rightarrow l_L^c + \phi^\dagger)$ due to the complex phases of
the Yukawa coupling matrix $h_{\alpha \beta}$, and a net lepton
number arises from the interference terms of the tree-level and
one loop diagrams \cite{luty,Flanz}.

In our leptogenesis scenario we have $\Gamma(N_R \rightarrow l_L + \phi) = \Gamma(N_R^c
\rightarrow l_L^c + \phi^\dagger)$ but there is a difference between the heavy light and left chirality neutrinos at thermal equilibrium due to the CP violating gravitational interaction (\ref{cpv0}),
\be
 n(N_R)-n(N_R^c)=\frac{T^2}{3} \frac{\dot R}{m_P M}\,.
\label{delta N}
\ee
The $N_R \leftrightarrow N_R^c$ interaction can be achieved by the scattering with a Higgs field. For example in SO(10) theories
Majorana masses for the right-handed neutrinos are generated either from
\be
{\bf 16_f\, 16_f \,{\overline {126}}_H} \supset y \,S^\prime\, N_R^c \,N_R\,,
\label{MNr1}
\ee
or from the non-renormalizable operators suppressed by some mass scale $\Lambda$
\be
 \frac{f}{\Lambda}{\bf 16_f\, 16_f \,{\overline{ 16}}_H \,{ \overline{ 16}}_H} \supset \frac{f}{\Lambda}\, S^2\, N_R^c\, N_R\,.
 \label{MNr2}
\ee
When the GUT Higgs fields $S^\prime$ or $S$ acquire a vev, a large Majorana  mass $M$ is generated for $N_R$ which breaks lepton number spontaneously. This following the see-saw mechanism leads to small neutrino masses at low energies. At temperatures larger than the heavy neutrinos and the GUT Higgs masses one there will be chirality flip scattering interactions like $S + N_R \leftrightarrow S+  N_R^c$ which change the lepton number. The interaction rate is given by
\be
\Gamma(S N_R\leftrightarrow S N_R^c)= \langle n_s \sigma \rangle= \frac{0.12}{\pi} \left(\frac{f}{\Lambda}\right)^2 T^3\,.
\ee
The interactions decouple at a temperature $T_D$ when $\Gamma(T_D)=H(T_D)$ from which we derive the decoupling temperature to be
\be
T_D=13.7 \pi \sqrt{g_*} \left(\frac{\Lambda}{f} \right)^2 \frac{1}{M_P}= 13.7 \pi \sqrt{g_*} \left(\frac{\langle S\rangle^2}{M} \right)^2 \frac{1}{m_P}\,,
\label{TD}
\ee
where we have used $M=f \langle S \rangle/\Lambda$.
The lepton number asymmetry is frozen in the heavy neutrino number asymmetry at temperature $T_D$ and is passed on to the light lepton sector after  the subsequent decays of $N$ into light neutrino. Substituting (\ref{TD}) in the expression for lepton asymmetry (\ref{delta N}) we obtain the value of frozen in lepton asymmetry as
\be
\eta= \frac{n(N_R)-n(N_R^c)}{s}= \frac{15}{2 \pi^2 g_*}\frac{\dot R}{T_D \,M \,m_P }\,,
\label{eta}
\ee
where $\dot R$ is to be evaluated at $T=T_D$. In the subsequent sections we first give a survey of $f(R)$ gravity and then compute $\dot R$ in $f(R)$ cosmology.

A comment is in order. In the case in which the fermion is, for example, an electron one also gets a splitting of energy levels
$E(e_R ) - E(e_L)$, but this does not lead to lepton generation of lepton asymmetry as both $e_L$ and $e_R$ carry the same lepton number.

\section{Fundamental equations and BBN in $f(R)$ theories of gravity}

As pointed out by Nojiri and Odintsov in \cite{nojiri}, modified gravity contains many topics which make these models very attractive, as for example:
1) they provide a natural unification of the early-time inflation and the later-time acceleration of the Universe owing to
the different role of the gravitational terms relevant at small and large scales; 2) they allow to unify dark matter and dark energy; 3)
they provide a framework for the explanation of hierarchy problem and unification of GUT with gravity. However, many $f(R)$ models of gravity
are strongly constrained (or ruled out) by solar system tests. In this respect, available models are those proposed in Ref. \cite{hu}
 \begin{equation}\label{hu}
    f(R)=-m^2\frac{c_1(R^2/m^2)^{2n}}{1+c_2(R/m^2)^{2n}}\,,
 \end{equation}
and in Ref. \cite{starob}
  \begin{equation}\label{starob}
    f(R)=R+\lambda R_{st}\left[\left(1+\frac{R^2}{R^2_{st}}\right)^{-d}-1\right]\,,
  \end{equation}
where $\lambda$, $m^2$, $c_{1, 2}$, $R_{st}$, $n$, and $d$ are free parameters.

These functions can be expanded in the appropriate regimes, reproducing simplest form of $f(R)$. A particular subclass
is given by
 \begin{equation}\label{f(R)=R+aR**n}
    f(R)=R+\alpha R^n\,,
 \end{equation}
where $\alpha>0$ has the dimensions [energy]$^{-2(n-1)}$ and $n>0$. Combinations of the free parameters allow to get a description of cosmic acceleration.
Particularly interesting is the case $n=2$ (referred in literature as Starobinsky's model \cite{starobinsky1980})
 \begin{equation}\label{f(R)=R+aR**2}
    f(R)=R+\alpha R^2\,.
 \end{equation}
The model (\ref{f(R)=R+aR**2}) has been studied in the framework
of astrophysics and cosmology. For instance, gravitational radiation emitted by isolated system constraints the free parameter
to $|\alpha| \lesssim (10^{17} - 10^{18})$m$^2$ \cite{berry,jetzer}. E{\"o}t-Wash experiments lead instead to the constraints
\begin{equation}\label{eot-washed}
    |\alpha|\lesssim 2 \times 10^{-9} \mbox{m}^2\,.
\end{equation}
However, more stringent constraints are provided by the Cosmic Microwave
Background (CMB) physics. The amplitude of the curvature perturbation corresponding to (\ref{f(R)=R+aR**2}) is $P_{\cal R}\simeq \displaystyle{\frac{N_k^2}{18\pi}\frac{1}{\alpha m_P^2}}$, with $N_k\sim 55$. Using the WMAP 5-years data ($P_{\cal R}\sim 2.445 \times 10^{-9}$) \cite{WMAP-5}, it follows that $\alpha$ is constrained as \cite{defelice}
 \begin{equation}\label{WMAP}
 |\alpha| < 10^{-39}{\mbox m}^2\,.
 \end{equation}
The bound (\ref{WMAP}) is obtained in the regime $R\gg \alpha^{-1}$ (in this regime the model describes the inflationary epoch).

Finally, we mention the model
 \begin{equation}\label{R**n-0}
    f(R)=R^{1+\epsilon}\,,
 \end{equation}
where $\epsilon \ll 1$. The tightest bound on $\epsilon$ has been inferred by Clifton and Barrow
\cite{barrow} in the framework of perihelion precession of Mercury
 \begin{equation}\label{epsilon}
    0 \leq \epsilon \lesssim 7.2 \times 10^{-19}\,.
 \end{equation}

The field equation obtained by the variation of the action (\ref{Lagr}) with respect to the
metric are
\begin{equation}\label{fieldeqs}
  f' R_{\mu\nu}-\frac{f}{2}\, g_{\mu\nu}-\nabla_\mu \nabla_\nu f'
  +g_{\mu\nu}\Box f'=\kappa^2 T_{g\, \mu\nu}\,,
\end{equation}
where the prime stands for the derivative with respect to $R$. The trace reads
\begin{equation}\label{tracef}
  3\Box f'+f' R-2f=\kappa^2 T_g\,,
\end{equation}
with $T_g$ the trace of the energy-momentum tensor.

In the spatially flat Friedman-Robertson-Walker (FRW) metric
\begin{equation}\label{FRWmetric}
 ds^2=- dt^2+a^2(t)[dx^2+dy^2+dz^2]\,,
\end{equation}
Eqs. (\ref{fieldeqs}) and (\ref{tracef}) become
\begin{eqnarray}
3f' H^2-\frac{R f'-f}{2}+3H f'' {\dot R}=\kappa^2 \rho\,, \label{0-0} \\
-2 f' H^2-f''' {\dot R}^2+f''(H f'' {\dot R}-{\ddot R})=\kappa^2 (\rho +p)\,, \label{i-j} \\
3f''' {\dot R}^2+3f'' {\ddot R}+9H f'' {\dot
R}+f'R-2f=\kappa^2 T_g\,, \label{trace}
\end{eqnarray}
where $H={\dot a}/a$, $T_{g\,0}^0=-\rho$, $T^i_{g\, j}=p\delta^i_j$, and $T_g=\rho-3p$. Moreover,
the Bianchi identities give a further condition on the
conservation of the energy
\begin{equation}\label{EnCons}
  {\dot \rho}+3\frac{\dot a}{a}(\rho +p)=0\,.
\end{equation}
In what follows, we shall look for those solutions of field equations such that the scale factor evolves
as
 \begin{equation}\label{a(t)}
a(t) = a_0 t^\beta\,, \qquad H=\frac{\beta}{t}\,.
 \end{equation}
The scalar curvature turns out to be
 \begin{equation}\label{R}
R=6 (2H^2+{\dot H}) = {\frac{6\beta(2\beta-1)}{t^2}}\,.
 \end{equation}
The $f(R)$ model we concern here is that one of Eq.  (\ref{f(R)=R+aR**n}).
By using Eqs. (\ref{0-0}) and (\ref{i-j}) and the usual expression relating the energy density and the pressure,
$p=w\rho$, where $w$ is the adiabatic index, one gets
 \begin{equation}\label{w-n}
    w=\frac{1}{3}+\delta(t)\,,
 \end{equation}
where
 \begin{equation}\label{delta}
    \delta\equiv \frac{2}{3\beta}\left(\frac{\beta+n {\cal A}}{\beta+{\cal A}}-2\beta\right)\,,
 \end{equation}
 with
 \[
 {\cal A}\equiv \alpha R^{n-1}[\beta(2-n)-(n-1)(2n-1)]\,.
 \]
The energy density $\rho$ assumes the form
 \begin{equation}\label{rho-n}
   \kappa^2 \rho=\frac{3\beta^2}{t^2}\left(1+\frac{\cal A}{\beta}\right)\,.
 \end{equation}
Notice that during the radiation dominated era ($\beta=1/2$), to which we are mainly interested, the quantity ${\cal A}$ vanishes
because $R=0$, as well as the perturbation $\delta$, and the adiabatic index reduces to the standard value $w=1/3$.

For our estimations on the baryon asymmetry, it is relevant to consider the following two features.

Let us first discuss the BBN constraints on $f(R)$ theories of gravity (see also \cite{barrow,kang}).
For the BBN scenario, one has to consider the weak interaction rate of particles ($p, n, e^{\pm}$ and $\nu$) in thermal equilibrium.
For $T\gg {\cal Q}$ (${\cal Q}=m_n-m_p$, where $m_{n,p}$ are the neutron and proton masses), one gets
$\Lambda(T)\simeq q T^5$, where $q=9.6 \times 10^{-46}\mbox{eV}^{-4}$ \cite{bernstein,kolb}.
To estimate the primordial mass fraction of ${}^4 He$, one defines $Y_p\equiv \lambda \, \frac{2 x(t_f)}{1+x(t_f)}$,
where $\lambda=e^{-(t_n-t_f)/\tau}$. $t_f$ and $t_n$ are the time of the freeze-out of the weak interactions and of the nucleosynthesis,
respectively, $\tau\simeq 887$sec is the neutron mean life, and $x(t_f)=e^{-{\cal Q}/T(t_f)}$ is the neutron to proton equilibrium ratio.
The function $\lambda(t_f)$ represents the fraction of neutrons that decay into protons in the time $t\in [t_f, t_n]$.
Deviations from $Y_p$ (generated by the variation of the freezing temperature $T_f$) are given by \cite{torres}
 \begin{equation}\label{deltaYp}
    \delta Y_p=Y_p\left[\left(1-\frac{Y_p}{2\lambda}\right)\ln\left(\frac{2\lambda}{Y_p}-1\right)-\frac{2t_f}{\tau}\right]
    \frac{\delta T_f}{T_f}\,.
 \end{equation}
In the above equation we have set $\delta T(t_n)=0$ because $T_n$ is fixed by the deuterium binding energy.
The current estimation on $Y_p$, $Y_p=0.2476\pm \delta Y_p$, with $|\delta Y_p| < 10^{-4}$ \cite{coc},
leads to
 \begin{equation}\label{deltaT/Tbound}
    \left|\frac{\delta T_f}{T_f}\right| < 4.7 \times 10^{-4}\,.
 \end{equation}
The freeze-out temperature $T$ is determined by $\Lambda= H$. One gets $T=T_f(1+\frac{\delta T_f}{T_f})$, where $T_f\sim 0.6$ MeV and
 \begin{equation}\label{deltaT/T}
    \frac{\delta T_f}{T_f} = \delta \frac{4\pi}{15}\sqrt{\frac{\pi g_*}{5}}\frac{1}{qm_P T_f^3}\simeq 1.0024 \left(\beta-\frac{1}{2}\right)\,.
 \end{equation}
Equations (\ref{deltaT/T}) and (\ref{deltaT/Tbound}) imply
 \begin{equation}\label{boubdc}
    2\beta-1 \lesssim 9.4 \times 10^{-4}\,.
 \end{equation}
As we shall see, the observed value of $\eta_{\mbox{exp}}$, Eq. (\ref{etaexp}), is obtained in our model of Leptogenesis
for values of $2\beta-1$ well below the bound (\ref{boubdc}).

The second aspect we wish to discuss concern the regime for which the terms that modify general relativity induce very tiny deviations from the standard cosmological model, i.e.
 \begin{equation}\label{alphaR}
 \alpha R^{n-1} \lesssim 1\,.
 \end{equation}
The latter implies that the parameter $\alpha$ is bound from above by
 \begin{equation}\label{alpha-n}
    \frac{\alpha}{(\mbox{m}^2)^{n-1}} \lesssim \Pi\,,
 \end{equation}
where
 \begin{equation}\label{Pi-n}
 \Pi \equiv \frac{1}{[6\beta|2\beta-1|]^{n-1}}\left(\frac{45}{16\pi^3 g_*}\right)^{n-1}
  \frac{10^{-68(n-1)}}{5^{2(n-1)}}\left(\frac{m_P}{T}\right)^{4(n-1)}\,.
 \end{equation}

In this approximation (${\cal A}\ll \beta$) one also obtains

 \begin{equation}\label{delta-reg}
    \delta \simeq \frac{2}{3\beta}\left(1-2\beta+\frac{(n-1) {\cal A}}{\beta}\right) \ll 1\,.
 \end{equation}

\section{Gravitational leptogenesis and $f(R)$ gravity}

As already noted, in the standard cosmological model ${\dot R}$ vanishes during the radiation era. However, tiny deviations
from General Relativity give rise to a Ricci curvature,
as well as its first time derivative, different from zero so that a net lepton asymmetry can be generated.
By making use of the definition of Ricci scalar curvature (\ref{R}),  it follows
\begin{equation}
 {\dot R}=-\frac{12\beta(2\beta-1)}{t^3}\,.
\end{equation}
Using the relation between the cosmic time and the temperature
$\displaystyle{\frac{1}{t}}\simeq \displaystyle{(\frac{32\pi^3 g_*}{90})^{1/2}\frac{T^2}{m_{P}}}+{\cal O}\left(\frac{\cal A}{\beta}\right)$,
the parameter characterizing the lepton asymmetry (\ref{eta}) assumes the form (to leading order in $(2\beta-1)$)
\begin{eqnarray}
\eta &=& \frac{128 \pi^2}{3 \sqrt{5}} \beta (2\beta-1) \sqrt{\pi g_*}
\frac{T_D^5}{m_P^4 M}\simeq  \label{eta-n}
\end{eqnarray}
 \[
 \simeq (2\beta-1) 3.4 \times 10^{-10}\, \frac{10^{12}\mbox{GeV}}{M}\left(\frac{T_D}{10^{15}\mbox{GeV}}\right)^5\,.
 \]
In this equation $T_D$ is the decoupling temperature at which the lepton number violating GUT reactions which change the number of $N$ and $N^c$ go out of equilibrium. This lepton asymmetry can be converted into baryon asymmetry by the action of sphalerons in the electroweak era
\cite{rubakov}.

An inspection of (\ref{eta-n}) immediately revels that the observed baryon asymmetry can be obtained for $T_D\sim 10^{16}$GeV, $M\sim 10^{9}$GeV,
provided that $2\beta-1\simeq 3\times 10^{-9}$.

Another possibility to get $\eta\sim 10^{-10}$ is given by taking, for example,  $\{T_D\sim 10^{17}$GeV; $M=10^{12}$GeV, $2\beta-1\simeq 10^{-15}-10^{-13}\}$
or $\{T_D\sim 10^{17}$GeV; $M=10^9$GeV, $2\beta-1\simeq 10^{-18}-10^{-16}\}$. The value of the heavy neutrino mass $\sim 10^{12}$GeV is consistent with the atmospheric neutrino scale $m_\nu=0.05$ eV, obtained from the see-saw relation $m_\nu=m_D^2/M$ with the Dirac mass scale $m_D\sim {\cal O}(10)$ GeV. Similar conclusions holds also for $M=10^9$GeV.

The lepton asymmetry generated via (\ref{eta-n}) is passed on to the light neutrino sector when the heavy neutrino decays at temperature $T\sim M\sim (10^{9}-10^{12})$GeV. The effects of washed out are avoided by
considering the effective (five dimensional) operator violating the lepton number $\Delta L=2$, ${\cal{L}}_W=\displaystyle{\frac{C}{2M}}(\overline{ {l_{L}}^c }~ \tilde{\phi^*})(\tilde{\phi^\dagger}~ l_{L}) +  h.c.$. We suppress the generation indices. $l_{L}=(\nu, e^-)_L^T$ is the left-handed lepton
doublet, $\phi$ is the Higgs doublet, and $M$ corresponds in general to some large mass scale (identified in our case with the heavy neutrino mass) \cite{weinberg}.
The interaction rate  is $\displaystyle{\Gamma(\nu_{L} +\phi \leftrightarrow \nu_{R} +\phi) =\langle n_\phi~ \sigma\rangle =
\frac{0.122}{\pi} \frac{|C|^2 T^3}{M^2}}$.
In the electroweak era, when the Higgs field in ${\cal{L}}_W$
acquires a $vev$, $\langle\phi\rangle =v=174$GeV, the five dimensional Weinberg operator gives rise to
a neutrino mass matrix $m_{\nu}= \frac{v^2 C}{M}$.
The lepton number violating  interactions decouple
when $\Gamma(T_l)=H(T_l) $, which implies
 \begin{equation}\label{decouplingTD}
 T_l= (2\beta)   \frac{13.68 \pi g_*^{1/2} v^4}{m_\nu^2 \, m_{P}}\simeq 2\times 10^{14}\left(\frac{0.05\mbox{eV}}{m_\nu}\right)^2\mbox{GeV}\,,
 \end{equation}
Therefore the heavy neutrino decays occur at a temperature $T\simeq M \simeq (10^{9}-10^{12})$ GeV well below the temperature $T_l=2 \times 10^{14}$GeV at which the light-neutrino lepton number violating interactions are effective. As a consequence, the lepton number asymmetry from the decay of asymmetric number of heavy neutrino decays is not washed out by Higgs scattering with light neutrinos.

\subsection{The general case $n\neq 0$}

Here we discuss the model (\ref{f(R)=R+aR**n}) with generic $n$.
In Table I is reported the estimation of $\Pi$ given by (\ref{Pi-n}) for different values of  $n$, $T_D$ and $2\beta-1$.
We see that $\Pi \ll 1$ for $n>1$.
The values of the decoupling temperature $T_D$ and $2\beta-1$ are those before used for inferring the correct estimation of  $\eta$.
By making use of (\ref{alphaR}) one can properly fix $\alpha$ in order that our approximation holds.

The case $n<1$ deserves also a discussion. As shown in \cite{CapozzielloTsujikawa}, this case yields the constraint $n< 10^{-10}$ if considered as a available candidate for dark matter (the authors assume $\alpha<0$). Such a bound is obtained from the combination of solar system experiments with the the equivalence principle violation experiments. Using (\ref{Pi-n}) with $T_D=10^{16}$GeV and $2\beta-1\simeq 10^{-22}$, one obtains $\Pi\simeq  10^{36}$. In order that our approximation work, it is requires a parameter $\alpha$ bounded at least by $\alpha \lesssim 10^{-36}$(m$^2$)$^{n-1}$.
Also in this model a net lepton asymmetry of the order of $\sim 10^{-10}$ can be therefore obtained.

We do not investigate the case $n < 0$ because these $f(R)$ models of gravity are affected by instability problems \cite{faraoniRMP,faraonibook,defelice}.

\begin{table}[t]
\caption{\label{TableI}Estimations of $\Pi$ for varying $n$. The values of $T_D$ and $2\beta-1$
are those leading to $\eta\simeq 10^{-10}$. We fix $T_D=10^{16}$GeV (similar results follow for $T_D=10^{17}$GeV).
On the left are reported the values for $2\beta-1=10^{-9}$, on the right for $2\beta-1=10^{-15}$.}
\begin{tabular}{ccccc}
  $n$ & $\Pi$ & $\Big|$ & $n$ & $\Pi$   \\  \hline
 1.1 & $10^{-6}$ & $\Big|$ & 1.1 & $10^{-6}$ \\ \hline
  1.5 & $10^{-27}$ & $\Big|$ & $1.5$ & $10^{-24}$ \\ \hline
 3  & $10^{-104}$ & $\Big|$ & 3 & $10^{-92}$ \\
\end{tabular}
\end{table}

\subsection{The case $n=2$}

A particular subclass of the model studied in the previous Section is the case in which $n=2$. The $f(R)$ function is given by (\ref{f(R)=R+aR**2}).
The parameter $\alpha$ now has the dimensions m$^2$ ([energy]$^{-2})$. As already pointed out in the Introduction,
E{\"o}t-Wash experiments provide stringent constraints on $\alpha$ given by (\ref{eot-washed}).
On the other hand, more stringent bound can be inferred from CMB physics (\ref{WMAP}).

For $n=2$, the function $\Pi$ reads
 \[
 \Pi=1.8\times 10^{-61}\, \frac{1}{2\beta-1}\left(\frac{10^{16}\mbox{GeV}}{T}\right)^4\,.
 \]
Taking $2\beta-1=10^{-9}$ and $T_D=10^{16}$GeV one gets $\Pi\simeq 10^{-53}$, while for $2\beta-1=10^{-18}$ and $T_D=10^{17}$GeV it follows $\Pi\simeq 10^{-47}$.
These values are well below the upper bounds (\ref{eot-washed}) and (\ref{WMAP}). In both cases, as before discussed, the observed baryon asymmetry is obtained.

\section{The model $f(R)=R^{1+\epsilon}$}

In this Section we finally study the model
  \begin{equation}\label{f(R)}
 f(R)=\left(\frac{R}{A}\right)^{1+\epsilon}\,,
 \end{equation}
where $A$ is a constant with dimensions $m_P^{2\epsilon/(1+\epsilon)}$. Eqs (\ref{a(t)}),
(\ref{0-0}) and (\ref{i-j}) imply
\begin{equation}\label{n=2alpha}
  1+\epsilon=2\beta\,.
\end{equation}
The constraint on the parameter $\epsilon$ given by (\ref{epsilon}) and
Eq. (\ref{n=2alpha}) yield \cite{barrow}
\begin{equation}\label{epsislon-bound}
    0 \leq 2\beta-1 < 7.2 \times 10^{-19}\,.
\end{equation}
Taking $2\beta-1\simeq 7\times 10^{-19}$, $M\sim 10^{8}$GeV \cite{strumia} and $T_D\sim 5 \times 10^{17}$GeV, the lepton asymmetry given by Eq. (\ref{eta-n})
yields $\eta_L\sim 10^{-10}$. These results show that this model of extended theory of gravity can work for our mechanism.

\section{Conclusions}

In this paper we have shown that $f(R)$ models of gravity, described by (\ref{f(R)=R+aR**n}),
(\ref{f(R)=R+aR**2}), and (\ref{R**n-0}), are available frameworks for the generation of a net baryon asymmetry.
These models, besides to be able to give a dynamical description of dark energy and to be also favorite candidate for
the dark matter puzzle, are consistent with solar system experiments.

The matter-antimatter asymmetry scenario discussed in this paper relies on a CP violating gravitational interaction between the heavy Majorana neutrinos and the
Ricci curvature. The lepton asymmetry generated in this way is in thermal equilibrium during the radiation era at GUT scale.
The subsequent decays of these heavy neutrinos into the light standard model particles and the conversion (via sphaleron processes) of lepton asymmetry into baryon asymmetry can explain the observed baryon asymmetry of the universe. As we have shown, the leptogenesis mechanism works for decoupling temperatures of the order of GUT scale, $T_D\sim (10^{16}-10^{17})$GeV, heavy neutrino mass $M\sim (10^{8}-10^{12})$GeV, and finally for very tiny deviations from the standard evolution of the Universe during the radiation era, i.e. $a\sim t^\beta$ with $\beta\lesssim 1/2+10^{-9}$
(a bound below the upper bound obtained from BBN $2\beta-1 < 10^{-3}$).


\end{document}